\documentstyle{article}

\newcommand{\beq}[1]{\begin{equation} \label{#1} }
\newcommand{\eeq}    {\end{equation}}
\newcommand{\tst}{\textstyle \mathstrut}
\newcommand{\ve}{\varepsilon}
\newcommand{\oh}{\frac{1}{2}}

\begin{document}

\title{Muon capture by oriented nuclei: new possibilities\\
for studying induced pseudoscalar interaction}
\author{A.L.Barabanov,\enskip Yu.V.Gaponov,\enskip
B.V.Danilin,\enskip N.B.Shul'gina\\
{\it The Kurchatov Institute, 123182 Moscow, Russia}}
\date{\mbox{}}
\maketitle

\begin{abstract}
Angular distribution of neutrinos (recoil nucleus) in muon capture
for an allowed Gamov-Teller transition is considered by taking
account of hyperfine effects. This angular distribution is shown
to include a correlation of the form $\sim P_2(\cos{\theta})$,
where $\theta$ is the angle between the neutrino momentum and the
axis specifying the orientation of the initial mesic atom This
correlation, which arises only if the initial mesic atom is
aligned, proves highly sensitive to the form factor $g_P$ of
induced pseudoscalar interaction. The proposed method for
determinating $g_P$ may be realized for the transition $1^+\to
2^+$ from the ground state of the $^6$Li nucleus to the narrow
resonance of the $^6$He nucleus in continuum, as well as in the
processes like
$^{10}{\rm B}(3^+)\to ^{10}{\rm Be}(2^+)$,
$^{11}{\rm B}(3/2^-)\to ^{11}{\rm Be}(1/2^-)$,
$^{23}{\rm Na}(3/2^+)\to {}$
$^{23}{\rm Ne}(1/2^+)$ with transitions to bound states.
\end{abstract}

\section{Introduction}
\label{s1}

In the weak nucleon-lepton Hamiltonian, the term that is
associated with induced pseudoscalar interaction has received the
less study. From the measured rate of muon capture by a free
proton, the form factor of this interaction was estimated at
$g_P=8.7\pm 1.9$, which complies with a value of $8.4\pm 0.9$
predicted on the basis of the PCAC hypothesis
\cite{Mukhopadhyay_93}. However, experiments devoted to muon
capture by $^6$Li \cite{Deutsch_68} and $^{28}$Si
\cite{Brudanin_95} nuclei yield substantially lower values
of~$g_P$.

In the nonrelativistic limit, the constant $g_P$ and the constant
$g_M$ of weak magnetism enter into the Hamiltonian in the first
order in the ratio of the momentum transfer to the nucleon mass
$M$. This ratio is of order $\sim E_{\nu}/Mc^2$, where $E_{\nu}$
is the neutrino energy, which is about $\sim 10^{-3}$ in
$\beta$-decay and about $\sim 10^{-1}$ in muon capture. Obviously,
this ratio controls the sensitivity of the total probabilities of
the above processes to $g_P$ . A detailed analysis of results
obtained in \cite{Deutsch_68} for the probability of muon capture
by the $^6$Li nucleus was carried out in \cite{Shulgina_93} within
three-body $\alpha+2{\rm N}$ model. This analysis was based on the
method of hyperspherical functions and employed the expressions
derived in \cite{Balashov_78} for the probability of muon capture.
In accord with previous calculations, it was found that the
experimental probability of the Gamow-Teller transition from the
ground $1^+$ state of the $^6$Li nucleus to the ground $0^+$ state
of the recoil nucleus $^6$He is reproduced with the form factor
$g_P$ close to zero.

However, the sensitivity of the muon-capture probability to $g_P$
is not very high (see above). This brings up the question of
whether there are observables in muon capture that are more
sensitive to the constant $g_P$. In particular, the ratio of the
probabilities $w^+/w^-$ of muon capture from states of hyperfine
structure with angular momenta $F^{\pm}=J_i\pm 1/2$ is intensively
discussed at present as a candidate for such a role
\cite{Deutsch_93}-\cite{Kuzmin_94}. We sought the answer to this
question along different lines, considering the angular
distribution of neutrinos (recoil nucleus) in muon capture for an
arbitrary Gamow-Teller transition $J_i \to J_f=J_i\pm 1$,
$\pi_i=\pi_f$. We performed our analysis in the approximation of
only single allowed $s$-wave matrix element being taken into
account. In the same approximation, the angular distribution of
neutrinos was earlier considered in~\cite{Hwang_78} for the
particular case of the $1^+\to 0^+$ Gamow-Teller transition.

In calculating the angular distribution of neutrinos in muon
capture by a nucleus with a nonzero spin $J_i$, we must bear in
mind that, when a muon proves to be in the \mbox{$1s$-state} from
which capture occurs, the state of the nucleus-muon system is a
superposition of two states of the hyperfine structure with
angular momenta $F^{\pm}=J_i\pm 1/2$. Because of hyperfine
interaction, these states are split --- that is, they have
different energies $E^{\pm}$. The period
$\sim 2\pi\hbar/(E^+-E^-)$ of oscillations developed by the
relative phase of the split states is much less than the
characteristic time of muon capture. As a result the interference
of states with different angular momenta $F$ make no contribution
to the differential probability of muon capture. Therefore, the
angular distribution of neutrinos is determined by the sum
\beq{1.1}
\frac{dw({\bf n}_{\nu})}{d\Omega_{\nu}}=
\sum_FP(F)
\frac{dw^F({\bf n}_{\nu})}{d\Omega_{\nu}},
\eeq
where $P(F)$ is the probability that the state $|F>$ is populated.
Thus, irrespective of whether the processes of capture from states
with different $F$ are detected experimentally or not, the angular
distribution of neutrinos must be calculated separately for each
state of the hyperfine structure.

It should be emphasized that we mean here the differential
probability, involving effects of angular correlations. The total
probability contains no contributions from the interference of
states with different $F$. Therefore, the total probability of
muon capture can be calculated without taking hyperfine splitting
into account. Hence, it is given by
\beq{1.2}
w=\sum_FP(F)w^F.
\eeq

This paper is organized as follows. In Section \ref{s2}, we
consider the polarization features of mesic atoms. Conditions
under which the rank-2 spin-tensor of a mesic atom (in other
words, the alignment parameter) is nonzero are analyzed. In
Section \ref{s3}, we derive a general expression for the
differential probability of muon capture, including effects of
angular correlations and taking into account the hyperfine
structure of the initial mesic atom. In Section \ref{s4}, we
present the formula for the angular distribution of neutrinos for
muon capture from a given state of the hyperfine structure. In
this angular distribution, special attention is paid to the term
that is proportional to the alignment parameter of the mesic atom.
This term exhibits the highest sensitivity to the form factor
$g_P$. The principal results of this study are summarized in the
Conclusion.

\section{Polarization and alignment of mesic atoms}
\label{s2}

As was mentioned above, a muon is captured by a nucleus from the
$1s$-state of a mesic atom. In this state, a muon occurs after
a series of transitions. In intermediate states, fine and
hyperfine interactions result in a partial transfer of muon
polarization to the electron shell and nucleus \cite{Balashov_78}.
It is very difficult to perform accurate calculations for this
process. However, this is unnecessary from the viewpoint of
practical applications. The residual muon polarization in a given
state of the hyperfine structure of the $1s$-orbit is reliably
determined experimentally from the asymmetry in the electron
emission from muon decays.

When the angular momentum $F$ of a state of the hyperfine
structure is higher than $1/2$, mesic atoms may display
spin-orientation types that are more complicated than mere
polarization. Generally, the atom is described by the
superposition
\beq{2.1}
\Psi_F=\sum_{\xi} a_{\xi} (F)\Psi_{F{\xi}}
\eeq
of states with given projections ${\xi}$ of the angular momentum
$F$ onto the $z$~axis. By performing averaging over the ensemble
of atoms, we obtain the density matrix
\beq{2.2}
\rho_{{\xi}{\xi}'}(F)=\overline{a_{\xi}(F)a^*_{{\xi}'}(F)},
\qquad
\sum_{\xi}\rho_{{\xi}{\xi}}(F)=1.
\eeq
To describe the polarization of atoms, it is convenient to use the
spin-tensors
\beq{2.3}
\tau_{Qq}(F)=
\sum_{{\xi}{\xi}'}C^{F{\xi}'}_{F{\xi}Qq}\rho_{{\xi}{\xi}'}(F),
\qquad
\tau_{00}(F)=1.
\eeq
In the case of axially symmetric orientation, we have
\beq{2.4}
\tau_{Qq}(F)=\tau_{Q0}(F)\delta_{q0}.
\eeq

The rank-1 and rank-2 spin-tensors are given by
\beq{2.5}
\tau_{10}(F)=\frac{<{\xi}>}{(F(F+1))^{1/2}}, \qquad
\tau_{20}(F)=\frac{3<{\xi}^2>-F(F+1)}
{((2F-1)F(F+1)(2F+3))^{1/2}},
\eeq
where $<{\xi}^n>=\sum_{\xi}{\xi}^n\rho_{{\xi}{\xi}}(F)$.
The rank-1 spin-tensor is proportional to the polarization
\mbox{$p_1(F)=<{\xi}>/F$}, which is defined in such a way that
$p_1(F)=1$ only if the state corresponding to the maximum
projection ${\xi}=F$ is populated. It is convenient to introduce
the parameters of orientation of an arbitrary rank $Q$ that are
normalized by the same condition. We specify them as
\beq{2.6}
p_Q(F)=
\frac{\tau_{Q0}(F)}{\mathop{(\tau_{Q0}(F))}
\nolimits_{\max}}, \qquad
\mathop{(\tau_{Q0}(F))}\nolimits_{\max}
=C^{FF}_{FFQ0}.
\eeq
For Q = 1 and 2, we then obtain
\beq{2.7}
p_1(F)=\left(\frac{F+1}{F}\right)^{1/2}\tau_{10}(F),
\qquad p_2(F)=\left(\frac{(F+1)(2F+3)}
{F(2F-1)}\right)^{1/2}\tau_{20}(F).
\eeq

The density matrix of the ensemble of nonoriented atoms is
proportional to identity matrix; hence, we have
\beq{2.8}
\tau_{Qq}(F)=\delta_{Q0}\delta_{q0}.
\eeq
Accordingly, the parameters $\tau_{Q0}(F)$ or $p_Q(F)$
characterize the orientation of atoms for $Q\ge 1$. Under
realistic conditions of population, the
quantities~$\rho_{{\xi}{\xi}}(F)$ smoothly depend on
projections~${\xi}$. Therefore, the spin-tensors~$\tau_{Q0}(F)$ or
the parameters~$p_Q(F)$ decrease fast with increasing $Q$. For
this reason, the parameter $p_2(F)$ is rarely taken into account
when the polarization $p_1(F)$ is nonzero. However, if atoms are
oriented in such a way that the populations $\rho_{{\xi}{\xi}}(F)$
of magnetic substates are independent of the sign of ${\xi}$, all
odd moments $<{\xi}>,\,<{\xi}^3>\ldots$ vanish. In this case,
$p_2(F)$ is the first nonzero parameter of orientation. An
ensemble of atoms oriented in this way is referred to as an
aligned ensemble. Accordingly, $p_2(F)$ is called the alignment
parameter. It is can easily be seen that only atoms with angular
momenta $F>1/2$ can be aligned. It should be emphasized that the
parameter $p_2(F)$ is nonzero not only when atoms are aligned but
also when they are polarized. In accordance with the above
definition, $p_2(F)=1$ in the case of complete polarization.

In capture of fast muons by atoms, some degree of alignment can
appear owing to orbital angular momenta orthogonal to the
collision axis \cite{Balashov_78}. Preorientation of nuclei
capturing muons also results in alignment of mesic atoms. Indeed,
let a muon with polarization $p_1(s)=\sqrt{3}\tau_{10}(s)$
directed along the unit vector ${\bf n}_{\mu}$ proves to be in the
$1s$-state. Suppose that the nuclear orientation at this instant
is determined by the spin-tensors $\tau_{N0}(I)$ in the reference
frame with the $z$ axis directed along the same vector
${\bf n}_{\mu}$. Expanding the direct product of the nuclear and
muon wave functions in the functions of states of the hyperfine
structure as
\beq{2.9}
\sum_ma_m(I)\psi_{Im}
\sum_{\sigma}a_{\sigma}(s)\psi_{s\sigma}=
\sum_{F{\xi}}a'_{\xi}(F)\sum_{m\sigma}
C^{F{\xi}}_{Ims\sigma}\psi_{Im}\psi_{s\sigma},
\eeq
we find that the coefficients in this expansion are given by
\beq{2.10}
a'_{\xi}(F)=\sum_{m\sigma}
C^{F{\xi}}_{Ims\sigma}a_m(I)a_{\sigma}(s),\qquad
\sum_{F{\xi}}|a'_{\xi}(F)|^2=1.
\eeq
The quantity $P(F)=\sum_{\xi}|a'_{\xi}(F)|^2$ is the probability
of finding the mesic atom in the state with angular momentum $F$.
Taking the square of the equation relating the amplitudes and
performing averaging over ensembles, we arrive at a relation
between the spin density matrices of nuclei, muons, and mesic
atoms. Going over to the spin-tensors, we obtain
\begin{eqnarray}
\lefteqn{\tau'_{Q0}(F)=
\left(\frac{(2F+1)^3}{(2Q+1)(2s+1)(2I+1)}\right)^{1/2}
\times{}}\nonumber\\[\medskipamount]
&&{}\times\sum_{NK}(2N+1)(2K+1)\tau_{N0}(I)\tau_{K0}(s)
C^{Q0}_{N0K0}\left\{
\begin{array}{ccc}
F&I&s\\
F&I&s\\
Q&N&K
\end{array}\right\},
\label{2.11}
\end{eqnarray}
where $\tau'_{00}(F)=P(F)$, and the symbol $\{\ldots\}$ denotes a
9j-symbol. The normalized spin-tensors of an ensemble of mesic
atoms with given $F$ can be written as
\beq{2.12}
\tau_{Q0}(F)=\frac{\tau'_{Q0}(F)}{\tau'_{00}(F)},\qquad
\tau_{00}(F)=1.
\eeq

If nuclei are not oriented ($\tau_{N0}(I)=\delta_{N0}$),
the spin-tensors of mesic atoms are given by
\beq{2.13}
\tau'_{Q0}(F)=\frac{2F+1}{(2s+1)(2I+1)}U(IsFQ,Fs)\tau_{Q0}(s),
\eeq
where $U(abcd,ef)=((2e+1)(2f+1))^{1/2}W(abcd,ef)$ is the
normalized Racah function \cite{Jahn_51}. The populations of
states of the hyperfine structure then coincide with the
statistical weights
\beq{2.14}
P(F)=\frac{2F+1}{(2s+1)(2I+1)}.
\eeq
For $Q\ge 1$, the only nonzero parameter among $p_Q(F)$ is the
polarization
\beq{2.15}
p_1(F)=\left\{ \begin{array}{ll}
-\frac{\tst 1}{\tst 3}p_1(s),&
F=I-\frac{\tst 1}{\tst 2};\\[\bigskipamount]
\frac{\tst 2I+3}{\tst 3(2I+1)}p_1(s),&
F=I+\frac{\tst 1}{\tst 2}.
\end{array}\right.
\eeq
If $F=I-\oh$, mesic atoms possess a nonzero polarization only for
$I\ge 1$.

It can be seen that oriented nuclei are necessary for mesic atom
to be aligned. In this case, it is convenient to single out two
contributions in the mesic-atom spin-tensors of even ranks
$Q=0,2\ldots$
\begin{eqnarray}
\tau'_{Q0}(F)&=&\frac{2F+1}{(2s+1)(2I+1)}\left[
U(sIFQ,FI)\tau_{Q0}(I)+{}
\lefteqn{\phantom{\left\{
\begin{array}{ccc}
F&I&s\\
F&I&s\\
Q&N&1
\end{array}\right\}}}\right.
\nonumber\\[\medskipamount]
{}&+&3\left(\frac{(2F+1)(2s+1)(2I+1)}{2Q+1}\right)^{1/2}
\tau_{10}(s)\times{}
\nonumber\\[\medskipamount]
&&\left.{}\times\sum_{N=1,3\ldots}(2N+1)\tau_{N0}(I)
C^{Q0}_{N010}\left\{
\begin{array}{ccc}
F&I&s\\
F&I&s\\
Q&N&1
\end{array}\right\}\right].
\label{2.16}
\end{eqnarray}
Of these, the second vanishes for unpolarized muons. Because of
this second term, populations of states of the hyperfine structure
differ from the statistical populations
\beq{2.17}
P(F)=\left\{
\begin{array}{ll}
\frac{\tst I}{\tst 2I+1}(1-p_1(s)p_1(I)),&
F=I-\frac{\tst 1}{\tst 2};\\[\bigskipamount]
\frac{\tst I+1}{\tst 2I+1}
(1+\frac{\tst I}{\tst I+1}p_1(s)p_1(I)),&
F=I+\frac{\tst 1}{\tst 2}.
\end{array}\right.
\eeq
It is clear that this effect is insignificant if the polarization
of muons is small by the instant at which they prove to be in the
$1s$-state.

According to (\ref{2.16}), the alignment of mesic atoms is
determined by the alignment of nuclei, on one hand, and by the
fact that nuclear polarization is superimposed on the polarization
of muons, on the other hand. The relative sign of these two
effects and, hence, the magnitude of alignment depend on the
direction of nuclear polarization, provided that the direction of
muon polarization is fixed. Setting the level populations equal to
their statistical values and disregarding the parameters $p_N(I)$
of nuclear orientation for $N\ge 3$, we find that the alignment of
mesic atoms is given by
\beq{2.18}
p_2(F)=\left\{
\begin{array}{ll}
p_2(I)-\frac{\tst 2}{\tst 5}p_1(s)p_1(I),&
F=I-\frac{\tst 1}{\tst 2};\\[\bigskipamount]
\frac{\tst (2I-1)(I+2)}{\tst (2I+1)(I+1)}\left(
p_2(I)+\frac{\tst 2(2I+3)}{\tst 5(2I-1)}p_1(s)p_1(I)\right),&
F=I+\frac{\tst 1}{\tst 2}.
\end{array}\right.
\eeq
Thus, we see that muon capture by atoms whose nuclei were
preoriented --- as was the case in experiments reported in
\cite{Kuno_87,Newbury_91} --- must lead to the formation of
aligned mesic atoms. Hence, additional correlations are expected
to appear in the angular distribution of neutrinos (recoil nuclei)
originating from the decays of mesic atoms aligned as the result
of muon capture. Explicit expressions for these correlations are
presented in Section \ref{s4}.

\section{Muon capture from a state of the hyperfine structure}
\label{s3}

The Hamiltonian for the muon-capture problem can be taken in the
form (see, for example, \cite{Balashov_78,Eisenberg_70})
\beq{3.1}
\hat{H}_W=-\frac{G\cos{\theta_C}}{\sqrt{2}}
\gamma_4(l)\gamma_{\lambda}(l)(1+\gamma_5(l))\hat{\tau}_+(l)
\sum_{j=1}^A\delta({\bf r}_j-{\bf r}_l)
\Gamma_{\lambda}(j)\hat{\tau}_-(j).
\eeq
where $G$ is the weak-interaction coupling constant, $\theta_C$ is
the Cabibbo angle, and $\gamma_{\lambda}$ matrices are taken in
the pseudo-Euclidean metric. This Hamiltonian describes the
pointlike interaction of leptons $l$ with each of the $A$
intranuclear nucleons numbered by the index $j$. The raising
$\hat{\tau}_+(l)$ and lowering $\hat{\tau}_-(j)$ operators act in
the isospin space and transform a muon into a neutrino and a
proton into a neutron, respectively.

The operator of the weak nucleon current is given by
\beq{3.2}
\Gamma_{\lambda}=\gamma_4
\left(g_V(k^2)\gamma_{\lambda}+
\frac{\hbar g_M(k^2)}{2Mc}\sigma_{\lambda\rho}k_{\rho}-
g_A(k^2)\gamma_{\lambda}\gamma_5-
i\frac{\hbar g_P(k^2)}{m_{\mu}c}k_{\lambda}\gamma_5\right).
\eeq
It involves the matrices $\sigma_{\lambda\rho}=(\gamma_{\lambda}
\gamma_{\rho}-\gamma_{\rho}\gamma_{\lambda})/2i$, and the
4-momentum transfer
\beq{3.3}
\hbar k_{\lambda}=p_{\lambda}-n_{\lambda}=
\nu_{\lambda}-\mu_{\lambda},
\eeq
where $p_{\lambda}$, $n_{\lambda}$, $\mu_{\lambda}$ and
$\nu_{\lambda}$ are the 4-momenta of the proton, neutron, muon,
and neutrino, respectively. The operator of the weak nucleon
current also contains the form factors of vector interaction
$g_V$, axial-vector interaction $g_A$, weak magnetism $g_M$, and
induced pseudoscalar interaction $g_P$. These form factors depend
on $k^2=k_{\lambda}k_{\lambda}$.

The probability of muon capture is determined according to the
Fermi rule. In going over to the nonrelativistic description of
intranuclear nucleons, the Hamiltonian of muon capture is
subjected to the Foldy-Wouthuysen transformation (see, for
example, \cite{Eisenberg_70}). Following common practice, we use
the transformed Hamiltonian evaluated to first-order terms in
$1/M$. It should be noted that second-order corrections in $1/M$
must involve terms associated with nucleon-nucleon potentials.

Let $\psi_{\mu}({\bf r}_{\mu},\sigma_{\mu})$ be the 4-component
wave function that describes a muon in the $1s$-state with the
projection $\sigma_{\mu}$ of the spin $s_{\mu}=1/2$ onto the $z$
axis. Similarly, $u_{\nu}({\bf p}_{\nu},\sigma_{\nu})$ is a
4-spinor that describes a neutrino with momentum ${\bf p}_{\nu}$
and with the projection $\sigma_{\nu}$ of the spin $s_{\nu}=1/2$
onto the $z$ axis. Following \cite{Eisenberg_70}, we introduce the
4-vector
\beq{3.5}
{\rm B}_{\lambda}(\sigma_{\mu},\sigma_{\nu})=
i\;\overline{\psi_{\mu}({\bf r}_{\mu},\sigma_{\mu})}
\gamma_{\lambda}(1+\gamma_5)
u_{\nu}({\bf p}_{\nu},\sigma_{\nu}).
\eeq
In the nucleon space, the nonrelativistic Hamiltonian of muon
capture can then be represented as
\begin{eqnarray}
\lefteqn{\hat{h}(\sigma_{\mu},\sigma_{\nu})=
\frac{G\cos{\theta_C}}{\sqrt{2}}\sum_{j=1}^A
\exp{\left(-i\frac{{\bf p}_{\nu}{\bf r}_j}{\hbar}\right)}\times{}}
\nonumber\\[\medskipamount]
&&{}\times\left\{-i{\rm B}^+_4(\sigma_{\mu},\sigma_{\nu})\left[
g_V(1+\ve_{\nu})+
(g_P-g_A)\ve_{\nu}(\hat{\vec{\sigma}}_j{\bf n}_{\nu})+
g_A(\hat{\vec{\sigma}}_j\frac{\hat{\bf p}_j}{Mc})\right]+{}\right.
\nonumber\\[\medskipamount]
&&\left. {}+{\bf B}^+(\sigma_{\mu},\sigma_{\nu})\left[
g_A\hat{\vec{\sigma}}_j-i(g_V+g_M)\ve_{\nu}[\hat{\vec{\sigma}}_j
\times{\bf n}_{\nu}]+
g_V\frac{\hat{\bf p}_j}{Mc}\right]\right\}\hat{\tau}_-(j),
\label{3.6}
\end{eqnarray}
where $\ve_{\nu}=E_{\nu}/2Mc^2$, $E_{\nu}$ is the neutrino energy,
and ${\bf n}_{\nu}={\bf p}_{\nu}/p_{\nu}$ is the unit vector
directed along the neutrino momentum. Since weak nucleon-lepton
interaction is pointlike, the 4-vectors
${\rm B}_{\lambda}(\sigma_{\mu},\sigma_{\nu})$ appearing in the
Hamiltonian can be taken at ${\bf r}_{\mu}={\bf r}_j$. The
operators $\hat{\vec{\sigma}}_j$ and
$\hat{\bf p}_j=-i\hbar\partial/\partial{\bf r}_j$ act in the space
of the $j$-th nucleon. The nucleon coordinates ${\bf r}_j$ are
reckoned from the center of mass of the nucleus. The above
Hamiltonian is a matrix in the space of the projections
$\sigma_{\mu}$ and $\sigma_{\nu}$. The energy of the emitted
neutrino is fixed to be
\beq{3.7}
E_{\nu}=E_f\left[\left(1+
\frac{2Q_{\mu}}{E_f}\right)^{1/2}-1\right]\simeq
Q_{\mu}\left(1-\frac{Q_{\mu}}{2E_f}+\ldots\right).
\eeq
In this expression, $Q_{\mu}=E_i+E_{\mu}-E_f$, where $E_i$ and
$E_f$ are the total energies (including the rest masses) of the
initial and final nuclei, respectively, and $E_{\mu}$ is the total
muon energy, including its binding energy in the atom before
capture.

Let $|J_fM_f>$ be the wave function that describes the internal
state of the final nucleus with spin $J_f$ and its projection
$M_f$ onto the $z$ axis. At the same time, the initial state of
the system involving a nucleus with spin~$J_i$ and a muon with
total angular momentum~$F$ is represented as
\beq{3.8}
|F>=\sum_{\xi}a_{\xi}(F)\sum_{M_i\sigma_{\mu}}
C^{F{\xi}}_{J_iM_is_{\mu}\sigma_{\mu}}
|J_iM_i>\psi_{\mu}(\sigma_{\mu}).
\eeq
Substituting these functions into the Fermi rule, we find that the
differential probability of muon capture per unit time from a
given state $|F>$ of the hyperfine structure is given by
\begin{eqnarray}
\frac{dw^F({\bf n}_{\nu})}{d\Omega_{\nu}}&=&
\frac{1}{\left(2\pi\hbar^2\right)^2}
\sum_{\sigma_{\nu}M_f}
\left|\sum_{\xi}a_{\xi}(F)\sum_{M_i\sigma_{\mu}}
C^{F{\xi}}_{J_iM_is_{\mu}\sigma_{\mu}}
<J_fM_f|\hat{h}|J_iM_i>\right|^2\times{}
\nonumber\\[\medskipamount]
&&\phantom{\frac{1}{\left(2\pi\hbar^2\right)^2}
\sum_{\sigma_{\nu}M_f}\sum_{\xi}a_{\xi}(F)}
{}\times\frac{E_{\nu}^2}{c^3(1+E_{\nu}/E_f)}.
\label{3.9}
\end{eqnarray}

In the nonrelativistic approximation, the muon wave function has
the form
\beq{3.10}
\psi_{\mu}({\bf r}_{\mu},\sigma_{\mu})\simeq\psi_{1s}(r_{\mu})
\left(\varphi_{\mu}(\sigma_{\mu}) \atop 0\right),
\eeq
where $\varphi_{\mu}(\sigma_{\mu})$ is a conventional
two-component spinor. In Hamiltonian (\ref{3.6}), the spatial
function is taken at ${\bf r}_{\mu}={\bf r}_i$. Since this
function weakly varies over the nuclear volume, its mean value is
usually taken outside the matrix-element sign. The square of this
mean value is
\beq{3.11}
<\psi_{1s}>^2=R(Z)\psi^2_{1s}(0)=
\frac{R(Z)Z^3}{\pi}\left(\frac{e^2}{\hbar c}\right)^3
\left(\frac{m'_{\mu}c}{\hbar}\right)^3,
\eeq
where $Z$ is the charge of the initial nucleus, and
$m'_{\mu}=m_{\mu}/(1+m_{\mu}c^2/E_i)$ is the reduced mass of the
muon-nucleus system. With the aid of the correction factor $R(Z)$,
it is considered that the nucleus is not a pointlike particle. To
write the expression for the probability of muon capture per unit
time in more compact form, we introduce the notation
\beq{3.12}
A_{\mu}=\lambda_{\mu}
\frac{8R(Z)Z^3}{3}
\frac{(E_{\nu}/m_{\mu}c^2)^2}
{(1+E_{\nu}/E_f)(1+m_{\mu}c^2/E_i)^3},
\eeq
where
\beq{3.13}
\lambda_{\mu}=\left(\frac{e^2}{\hbar c}\right)^3
\frac{(G\cos{\theta_C})^2(m_{\mu}c^2)^5}
{{\strut \hbar}^7 {\strut c}^6}\simeq
1.005 \cdot 10^3 \quad {\mbox s}^{-1}.
\eeq

According to (\ref{3.6}) and (\ref{3.9}), the probability of muon
capture is determined by nuclear matrix elements of four types.
Each of these is conventionally represented as a multipole
expansion (see, for example, \cite{Balashov_78}). For the
Gamow-Teller matrix element, this expansion has the form
\begin{eqnarray}
\lefteqn{<J_fM_f|\sum_{j=1}^A
\exp{\left(-i\frac{{\bf p}_{\nu}{\bf r}_j}{\hbar}\right)}
\hat{\sigma}_{jq}\hat{\tau}_-(j)|J_iM_i>={}}
\nonumber\\[\medskipamount]
&&{}=\frac{(4\pi)^{3/2}}{\sqrt{3}}
\sum_{wm}(-1)^wY^*_{wm}({\bf n}_{\nu})
\sum_{uM}C^{uM}_{1qwm}
C^{J_fM_f}_{J_iM_iuM}[1wu],
\label{3.15}
\end{eqnarray}
The phase factors $i^w$, together with spherical harmonics
$Y_{wm}({\bf r}_j)$, are conveniently incorporated into the
reduced matrix elements $[1wu]$. Owing to this, the reduced matrix
elements are real-valued quantities, provided that the nuclear
wave functions are transformed under the time reversal in the
standard way \cite{Bohr_69}
\beq{3.18}
\hat{T}|JM>=(-1)^{J+M}|J-M>.
\eeq
The spherical components $\hat{\sigma}_{jq}$ of the vector
operator $\hat{\vec{\sigma}}_j$ are defined as
$\hat{\sigma}_{j\pm 1}=
\mp(\hat{\sigma}_{jx}\pm \hat{\sigma}_{jy})/\sqrt{2}$,
$\hat{\sigma}_{j0}=\hat{\sigma}_{jz}$.

\section{Angular distribution of neutrinos in a Gamow-Teller
transition}
\label{s4}

In this section, we derive the expression for the angular
distribution of neutrinos for muon capture by a nucleus with
nonzero spin $J_i$. In accordance with what was said in the
Introduction, we consider an ensemble of mesic atoms in a given
state of the hyperfine structure with angular momentum $F$. Let
the orientation of the ensemble be specified by the parameters
$p_Q(F)$. The axis of orientation ($z$ axis) is directed along the
unit vector ${\bf n}_{\mu}$. We consider muon capture accompanied
by the transition of the nucleus into the $J_f=J_i\pm 1$ state
without change in parity ($\pi_f=\pi_i$). We can then expect that
the Gamow-Teller matrix element is dominant. According to
multipole expansion (\ref{3.15}), we have
\beq{4.3}
<J_fM_f|\sum_{j=1}^A
\exp{\left(-i\frac{{\bf p}_{\nu}{\bf r}_j}{\hbar}\right)}
\hat{\sigma}_{jq}\hat{\tau}_-(j)|J_iM_i>\simeq
\frac{4\pi}{\sqrt{3}}
C^{J_fM_f}_{J_iM_i1q}[101],
\eeq
that is, we confine our treatment to a single allowed s-wave
matrix element. In this approximation, the calculation according
to formulas (\ref{3.6}) and (\ref{3.9}) shows the angular
distribution of neutrinos (recoil nuclei) can be represented as
\beq{4.4.1}
\frac{dw^F({\bf n}_{\nu})}{d\Omega_{\nu}}=
\frac{A_{\mu}[101]^2}{4\pi}\frac{2J_f+1}{2J_i+1}
\left(a_0+a_1p_1(F)({\bf n}_{\nu}{\bf n}_{\mu})+
a_2p_2(F)\frac{3({\bf n}_{\nu}{\bf n}_{\mu})^2-1}{2}\right),
\eeq
where
\begin{eqnarray}
a_0&=&\left(g^2_A-
\frac{2}{3}g_A(g_P-g_A)\ve_{\nu}-
\frac{4}{3}g_A(g_V+g_M)\ve_{\nu}\right)
C_1(J_i,J_f,F),\label{4.4.2}\\[\medskipamount]
a_1&=&-\left(g^2_A-2g_A(g_V+g_M)\ve_{\nu}\right)
C_2(J_i,J_f,F)-{}
\nonumber\\[\medskipamount]
&&{}-\left(g^2_A-g_A(g_P-g_A+g_V+g_M)\ve_{\nu}\right)
C_3(J_i,J_f,F),
\label{4.4.3}\\[\medskipamount]
a_2&=&g_A(g_P-g_A-g_V-g_M)\ve_{\nu}
C_4(J_i,J_f,F).\label{4.4.4}
\end{eqnarray}
This angular distribution is a series in the Legendre polynomials
$P_0(\cos\theta)=1$,
$P_1(\cos\theta)=\cos\theta$,
$P_2(\cos\theta)=3(\cos^2\theta -1)/2$, where $\theta$ is the
angle between the direction of neutrino momentum and the axis of
orientation of the initial mesic atom. Since the Hamiltonian of
muon capture was taken here in the approximation that is linear in
$\ve_{\nu}=E_{\nu}/2Mc^2$, we retained terms of the same order in
the expression for the angular distribution. The coefficients
$C_j(J_i,J_f,F)$ are given by
\beq{4.5}
C_1(J_i,J_f,F)=1+\sqrt{6}
U(F\oh J_i1,J_i\oh)U(J_f1J_i1,J_i1),
\eeq
\begin{eqnarray}
C_2(J_i,J_f,F)=
\left(\frac{F}{F+1}\right)^{1/2}\left[
\sqrt{2}U(\oh J_iF1,FJ_i)U(J_f1J_i1,J_i1)-{}\right.&&
\nonumber\\[\medskipamount]
\left.\phantom{C_2(J_i,J_f,F)=
\left(\frac{F}{F+1}\right)^{1/2}
\sqrt{2}U(\oh J_iF1)}
-\sqrt{3}U(J_iF\oh 1,\oh F)\right],&&
\label{4.6}
\end{eqnarray}
\begin{eqnarray}
&\lefteqn{C_3(J_i,J_f,F)=
2\left(\frac{F}{F+1}\right)^{1/2}\left[
\frac{1}{\sqrt{3}}U(J_iF\oh 1,\oh F)+{}
\phantom{\left\{
\begin{array}{ccc}
J_i&F&1/2 \\
J_i&F&1/2 \\
2  &1&1
\end{array}\right\}}
\right.}&
\nonumber\\[\medskipamount]
&&\left.{}+5(2(2J_i+1)(2F+1))^{1/2}
\left\{
\begin{array}{ccc}
J_i&F&1/2 \\
J_i&F&1/2 \\
2  &1&1
\end{array}\right\}
U(J_f1J_i2,J_i1)\right],
\label{4.7}
\end{eqnarray}
\begin{eqnarray}
C_4(J_i,J_f,F)=
2\left(\frac{10F(2F-1)}{(F+1)(2F+3)}\right)^{1/2}
\left[\frac{1}{3}U(\oh J_iF2,FJ_i)U(J_f1J_i2,
\lefteqn{J_i1)-{}
\phantom{\left\{
\begin{array}{ccc}
J_i&F&1/2 \\
J_i&F&1/2 \\
1  &2&1
\end{array}\right\}}}
\right.&&
\nonumber\\[\medskipamount]
\left.{}-((2J_i+1)(2F+1))^{1/2}
\left\{
\begin{array}{ccc}
J_i&F&1/2 \\
J_i&F&1/2 \\
1  &2&1
\end{array}\right\}
U(J_f1J_i1,J_i1)\right].
%\phantom{(2J_i+1)}
&&
\label{4.8}
\end{eqnarray}
The explicit expressions for these coefficients are presented in
the Appendix. In the particular case of $J_i=1 \to J_f=0$, the
formula that we derived for the angular distribution coincides
with that presented in \cite{Hwang_78} to the terms linear in
$\ve_{\nu}=E_{\nu}/2Mc^2$ that were taken into account.

It is interesting to note that each of the coefficients
$C_j(J_i,J_f,F)$ vanishes if the initial angular momentum $F$ of
the mesic atom and the final angular momentum $J_f$ of the nucleus
differ by $3/2$; that is,
\beq{4.9}
C_j(J_i,J_f=J_i\pm 1,F=J_i \mp\oh)=0,
\eeq
Obviously, we are dealing here with the trivial consequence of the
law of angular-momentum conservation. Indeed, in the approximation
of a single $s$-wave, the total angular momentum of the system
involving a final nucleus and a neutrino can be equal only to
$J_f\pm 1/2$. Therefore, transitions from states of the hyperfine
structure with $F=J_i\pm 1/2$ into nuclear states with
$J_f=J_i\mp 1$ are forbidden in the approximation that we use.

From expression (\ref{4.4.1}) for the angular distribution, we see
that the term proportional to $\sim P_2(\cos\theta)$, which
describes the anisotropy of neutrino emission in the directions
parallel and orthogonal to the vector ${\bf n}_{\mu}$ and which is
due to the alignment of the initial mesic atom, exhibits the
highest sensitivity to the form factor $g_P$ of induced
pseudoscalar interaction. This term is in direct proportion to the
combination $(g_P-g_A-g_V-g_M)$. The reason behind this
sensitivity is as follows. In our approximation of a single
$s$-wave in the exit channel, the dependence of the Hamiltonian on
the direction of ${\bf n}_{\nu}$is the only source of the angular
anisotropy of neutrino emission. The square of the matrix element
of Hamiltonian (\ref{3.6}) in turn depends on the vector
${\bf n}_{\nu}$
through two factors. First, the sums of the squared combinations
of the components of the 4-vector
${\rm B}_{\lambda}(\sigma_{\mu},\sigma_{\nu})$ over $\sigma_{\nu}$
are linear in ${\bf n}_{\nu}$. Second, Hamiltonian (\ref{3.6})
involves the product of the vector ${\bf n}_{\nu}$ and small
parameter $\ve_{\nu}$. If the angular distribution is computed to
first-order terms in $\ve_{\nu}$, terms that are quadratic in
${\bf n}_{\nu}$ must be in direct proportion to the difference
$g_P-g_A$ or to the sum $g_V+g_M$. But it is the result that we
obtained. From these speculations, it follows that correlations
that are quadratic in ${\bf n}_{\nu}$ must exhibit a similar
sensitivity to $g_P$ for mixed Fermi and Gamow-Teller transitions
as well.

\section{Conclusion}
\label{s6}

In this study, we have considered the angular distribution of
neutrinos (recoil nuclei) in muon capture for an arbitrary
Gamow-Teller transition, taking into account hyperfine effects. It
has been shown that, in the angular distribution, there is a
correlation proportional to $\sim P_2(\cos{\theta})$ (where
$\theta$ is the angle between the neutrino momentum and the axis
of orientation of the initial mesic atom), which is highly
sensitive to the form factor $g_P$ of induced pseudoscalar
interaction --- that is, this correlation is in direct proportion
to the combination $(g_P-g_A-g_V-g_M)$. However, such a
correlation appears provided that the initial mesic atom is
aligned. Among other things, this means that, for the $1^+\to 0^+$
transition in muon capture by the $^6$Li nucleus, the proposed
method for studying $g_P$ is unsuitable if the final nucleus in
the above transition is $^6$He in the ground state. At least, this
is so in the approximation used here, which takes into account the
single (leading) reduced matrix element $[101]$. Indeed, on one
hand, the state with angular momentum $F=1/2$ cannot be aligned.
On the other hand, muon capture from the state with $F=3/2$ is
forbidden in this approximation. However, this mechanism may prove
effective in studying muon capture by the same nucleus $^6$Li with
the transition into $J_f^{\pi}=2^+$ continuum resonance state of
the system involving an $\alpha$-particle
and two neutrons. The energy
and width of this state are $E=1.8$~MeV and $\Gamma=0.1$~MeV. The
neutrino momentum can be reconstructed by measuring the momenta of
three particles in the final state. In the case of the
$1^+\to 2^+$ transition, muon capture from the $F=1/2$ state of
the hyperfine structure, which cannot be aligned, is forbidden.
Recall that, here, we mean only the approximation of
the leading term associated with the reduced matrix element
$[101]$. At $J_i=1$, $J_f=2$ and $F=3/2$, we obtain
$C_1=3/2$, $C_2=-3/2$, $C_3=-3/5$, $C_4=2/5$ for the coefficients
appearing in the angular distribution.

Of course, the proposed method for determining $g_P$ from the
anisotropy of neutrino emission in the directions parallel and
orthogonal to the axis of orientation of mesic atoms is applicable
to any Gamow-Teller transitions. In particular, it is interesting
to investigate muon capture by $^{10}$B ($3^+\to 2^+$),
$^{11}$B ($3/2^-\to 1/2^-$),
$^{23}$Na ($3/2^+\to 1/2^+$), etc., nuclei that is accompanied by
transitions to bound states (these processes are discussed in
\cite{Deutsch_93,Gorringe_93}).
\bigskip

This work was supported by the International Science Foundation
(grant no. M7C300).
\bigskip\bigskip

\appendix

\noindent{\Large \bf Appendix}
\bigskip

\noindent By means of the explicit expressions for the Racah
functions \cite{Jahn_51} and for 9j-symbols \cite{Sobelman_63},
the coefficients $C_j(J_i,J_f,F)$ defined by
(\ref{4.5})--(\ref{4.8}) can be represented as follows
\medskip
$$
C_1=\left\{
\begin{array}{ll}
1+\frac{\tst J_i(J_i+1)+2-J_f(J_f+1)}{\tst 2J_i},&
F=J_i-\frac{\tst 1}{\tst 2};\\[\bigskipamount]
1-\frac{\tst J_i(J_i+1)+2-J_f(J_f+1)}{\tst 2(J_i+1)},&
F=J_i+\frac{\tst 1}{\tst 2};
\end{array}\right. \eqno (A1)
$$
\bigskip
$$
C_2=\left\{
\begin{array}{ll}
\frac{\tst (2J_i-1)
\left(J_i(J_i+3)+2-J_f(J_f+1)\right)}
{\tst 2J_i(2J_i+1)},&
F=J_i-\frac{\tst 1}{\tst 2};\\[\bigskipamount]
\frac{\tst J_i(J_i-1)-J_f(J_f+1)}{\tst 2(J_i+1)},&
F=J_i+\frac{\tst 1}{\tst 2};
\end{array}\right. \eqno (A2)
$$
\bigskip
$$
C_3=\left\{
\begin{array}{ll}
\lefteqn{\frac{\tst 1}{\tst J_i(2J_i+1)}\left[
\left(J_i(J_i+1)+1-J_f(J_f+1)\right)^2-{}\right.}&
\\[\bigskipamount]
\left.{}-J_i(3J_i+1)-
J_f(J_f+1)+1\right],&
F=J_i-\frac{\tst 1}{\tst 2};\\[\bigskipamount]
\lefteqn{\frac{1}{\tst (J_i+1)(2J_i+3)}\left[
J_i(3J_i+5)+J_f(J_f+1)+1-{}\right.}&
\\[\bigskipamount]
\left.{}-\left(J_i(J_i+1)+1-J_f(J_f+1)\right)^2\right],&
F=J_i+\frac{\tst 1}{\tst 2};
\end{array}\right. \eqno (A3)
$$
\bigskip
$$
C_4=\left\{
\begin{array}{ll}
\frac{\tst J_i-1}{\tst 3J_i(J_i+1)(2J_i+1)}\left[
\left(J_i(J_i+1)+2-J_f(J_f+1)\right)\times{}\right.
&\\[\bigskipamount]
\left.{}\times\left((J_i+1)(3J_i+2)-3J_f(J_f+1)\right)-
8J_i(J_i+1)\right],&
F=J_i-\frac{\tst 1}{\tst 2};\\[\bigskipamount]
\frac{\tst 1}{\tst 3(J_i+1)(2J_i+3)}\left[
\left(J_i(J_i+1)+2-J_f(J_f+1)\right)\times{}\right.
&\\[\bigskipamount]
\left.{}\times\left(J_i(3J_i+1)-3J_f(J_f+1)\right)-
8J_i(J_i+1)\right],&
F=J_i+\frac{\tst 1}{\tst 2}.
\end{array}\right. \eqno (A4)
$$

\end{document}